\documentstyle[12pt]{article}
\begin{document}
\def\strut{\rule[-.5cm]{0cm}{1cm}}
\def\dspace{\baselineskip = .30in}

\title{
\begin{flushright}
{\large\bf IFUP-TH 09/95}
\end{flushright}
\vspace{1.5cm}
\Large\bf INFLATION VERSUS THE COSMOLOGICAL MODULI PROBLEM}

\author{{\bf Gia  Dvali}\thanks{Permanent address: Institute of Physics,
Georgian Academy of Sciences,  \hspace{1cm}380077 Tbilisi, Georgia.
E-mail:
dvali@ibmth.difi.unipi.it }\\ Dipartimento di Fisica, Universita di Pisa
and INFN,\\ Sezione di Pisa I-56100 Pisa, Italy\\}

\date{ }
\maketitle

\begin{abstract} We show that in generic supergravity theories the mass of
the moduli during inflation is larger (or at least of the same order
of magnitude) than the Hubble constant. This fact does not depends on the
details of the inflation and on the value of the Hubble parameter  during it.
The reason is that inflationary universe is dominated by large $F$-term
(or $D$-term) density which is higher than the SUSY breaking scale in the
present minimum and stabilizes the flat directions of the
supersymmetric vacua.
 Therefore, in general even standard inflationary scenarios (with large $H$)
may solve the cosmological moduli problem.

\end{abstract}
\newpage

\dspace

 Supergravity and superstring theories often include scalar fields
with the weak scale ($M_W$) masses and Planck scale ($M_P$)
suppressed interactions.
Such fields have no supersymmetric mass term and/or renormalizable couplings
in the superpotential and in the supersymmetric limit they
parameterize the flat
directions in the SUSY vacuum. Their masses are induced by the same mechanism
that breaks supersymmetry and stabilizes flat directions. Since the message
about the SUSY breaking is carried by Planck scale suppressed interactions,
the resulting masses are of the order

\begin{equation}
 m^2 \sim { |F_i|^2 \over M_P^2}
\end{equation}

where $F_i$ are the expectation values of the F-terms that break supersymmetry.
Since the same mechanism is responsible for the transfer of SUSY breaking to
the visible (quark and lepton) sector the value of $F_i$ has to be of order
$M_WM_P$ resulting in the weak scale mass.
Independently from their origin, such fields having only Planck scale
couplings and weak scale (nonsupersymmetric) mass we will call
collectively as moduli fields and will denote them by $Z$.
It is well known that in the cosmological
context moduli exhibit serious problems. The cosmological evolution of the
moduli have been analyzed in great detail before [1].
The key point of the problem is that in the early
universe the expectation values of the moduli are expected to be
far (at distance $Z_0 \sim M_P$) from the minimum (due to thermal or quantum
fluctuations) and so one has a Bose condensate, which later starts oscillate
about the minimum and contributes to much energy density to the mass of
the universe. After late decay of this cold bosons the temperature of the
universe is too low for the standard nucleosynthesis.

 It is assumed usually [1-4], that it is very difficult to eliminate
moduli Bose
condensate (reduce $Z_0$) by inflation [5], unless during inflation the
Hubble
constant ($H$) is of order $M_W$ (mass of the moduli at the $minimum$).
(Alternative would be enormous number of e-foldings). Therefore in [3] the
inflation with weak scale $H$ was proposed as a possible (and perhaps the
only promising [4]) solution to the moduli problem. However, such late
inflation in general has problems in explaining the large density
fluctuations.

 In the present note we show that for eliminating the
initial moduli condensate by inflation such a small $H$ is not necessary,
because
moduli acquire large ($ \sim H$) mass during inflation. By definition
the moduli has a mass $\sim M_W$ ``today'' (in the true vacuum with
zero cosmological constant), but there was no reason for
this mass to be the same during inflation. In fact the opposite is
true: in the generic supergravity theories the mass of the moduli
during inflation is larger (or $\sim$) than the Hubble parameter.
This has to do with the fact that any inflation brakes supersymmetry [6],
since it provides a positive (and large) cosmological constant and thus
nonzero F-term density (which automatically gives large mass to moduli).
This mass is related to the value of the Hubble constant and as a matter
of fact is larger (or at least of the same order of magnitude).
This result is not surprising, since there is no generic reason why
the flat directions of SUSY vacua may survive in the state with
large vacuum energy and it is expecteble that corresponding zero modes
(moduli) will get masses of order the vacuum energy density ($\sim |F_i|^2$)
suppressed by the scale of messenger interaction ($M_P^2$).

To be more precise consider generic supergravity scalar potential:

\begin{equation}
 V = exp({K \over M^2})[K^{i-1}_jF_iF^{*j} - 3{|W|^2 \over M^2}]
\end{equation}

where $K(S_i,S_i^*)$ is a Kahler potential, $W(S_i)$ is superpotential
and $S_i$ are chiral superfields (their scalar components we denote
by the same symbols). $F_i$ -terms are given by
$F_i = W_i + {W \over M^2}K_i$ where upper (lower) index denotes derivative
with respect to $S_i$ ($S_i^*$) respectively and $M^2 = {M_P^2 \over 8\pi}$
is a reduced Planck mass. (We neglect possible $D$-terms and assume that
they vanish during inflation).

 Let us start with a simplest case in which moduli is decoupled from the
visible sector superfields both in Kahler potential and superpotential.
For definiteness let us denote moduli field by
$Z$ and assume simplest possible dependence:
\begin{equation}
K = |Z|^2 + K'(S_i,S_i^*)
\end{equation}
Where $K'$ and $W$ are arbitrary functions independent of $Z$. Now it is
easy to see that whatever the quantities $K', W$ are, in $any$ state
with zero or positive cosmological constant the only minimum (and even
extremum) in $Z$ is at $Z = 0$ and it's mass is given by:

\begin{equation}
m_Z^2 = {e^{K'/M^2} \over M^2} [K'^{i-1}_jF_iF^{*j} - 2 {|W|^2 \over M^2}]
\end{equation}

In particular, in the state with vanishing vacuum energy and SUSY breaking
scale
$F_i \sim M_WM$ this relation implies $m_Z \sim M_W$.

 Now let us consider how inflation affects this situation. Here we do not
want to advocate any particular inflationary scenario and/or discuss
whether such can be implemented in the supergravity framework. Our
aim is to point out some model-independent consequences of the inflation
if it happens in some way in above system.
 The basic idea of inflationary scenario [5] is that for some time
universe has to stay in the state with the positive (and large) cosmological
constant ($V > 0$) in order to undergo a period of exponential expansion.
This state may or may not be a local minimum of the potential, but essential
condition is that fields roll slowly enough and
the Hubble constant is given by

\begin{equation}
H^2 = {V \over 3 M^2}
\end{equation}

{}From above we have learned two important things about the moduli behavior
during (any) inflation:

1) Classical expectation value of $Z$ is trapped in the minimum at $Z = 0$;

2) As it can be easily seen from (2),(4) and (5), the mass of the moduli
is larger than the Hubble constant:
\begin{equation}
m_Z^2 = 3H^2 + {|W|^2 \over M^4} e^{K' \over M^2}
\end{equation}

Thus, there is no need to assume that $H \sim M_W$ since (6) is valid for
any $H$.
For the case $H < m_Z$ reduction of the initial expectation value $Z_0$
of moduli goes through the factor [3] $e^{{-3 \over 2} N}$ where $N$ is the
number of e-foldings since the beginning of inflation. Therefore,
we expect that any inflation which can be implemented in the supergravity
framework may dilute the coherent condensate and solve the moduli problem.

  Now let us consider the situation in which moduli has some Planck scale
suppressed couplings with the visible sector superfields in the
superpotential:

\begin{equation}
 W' = {1 \over M^{n-2}} Z S_1 S_2 ..S_n + ....
\end{equation}

Clearly, whatever the origin of the $S_i$ fields is, the following
condition should be satisfied:

(*) In the phenomenological minimum with zero cosmological constant
(present vacuum), in each n-linear invariant only $n - 4$ fields are
allowed to have large (Planck scale) VEVs.

Otherwise $Z$ will acquire large effective coupling and will not be
``moduli'' any more. In the case of GUT Higgs fields this requirement
can be restricted by operators up to $n = 11$ or so, since the highers
will be suppressed by large powers of $M_{GUT}/M$.

 Assuming that the minimum in $Z$ during inflation is still at $Z=0$,
we are lead to the following correction to the right-hand side of
(6)

\begin{equation}
\delta m_z^2 = e^{K' \over M^2} [|W'_{zz}|^2 + K'^{i-1}_j F_{iz}F^{*jz} -
{|W'_z|^2 \over M^2}]
\end{equation}

where lower (upper) index $z$ denotes derivative with respect to $z(z^*)$.
There are two model dependent possibilities:

a) If condition (*) is satisfied during inflation, then it is clear from
(8) that positive correction to (6) is at best of order $M_W^4/M^2$
and the negative one is $\sim M_W^6/M^4$ and therefore relation (6)
between $m_z$ and $H$ is valid with great accuracy.

b) In some models the condition (*) can be violated during inflation.
This may be the case if some $S$-fields in (7) are coupled to the inflaton
and get large ($\sim M$) displacement of the VEVs. In such a case the
corrections of the both sign in (8) can be of order $M^2$, but it would
be very unnatural to expect that they
cancel each other and (6) in such a way
that resulting mass of the moduli is of order $M_W$. It is even more
unnatural that such accidental ``fine tuning'' may hold along the
full inflationary trajectory. Therefore, we conclude that in general
moduli mass is of order (and in many cases larger) the Hubble constant
and therefore even standard inflation (with large $H$) can be
sufficient to solve the problem. Of course, in particular cases there
can be additional model-dependent factors (e.g. reheating temperature) which
may need some treatment and can not be discussed here.

 At the end let us ask what may happen if the minimum in $Z$ is displaced
(by $Z_0 \sim M$) from the origin during inflation? Again, this is
very model dependent situation, but due to above arguments we
expect that in this ``inflationary'' minimum the moduli mass again
will be $\sim H$. However, now this may not be enough to solve the
problem, since finally this displacement has to disappear. If this happens
while inflation is still going on and $m_z$ is large, problem may be
avoided. However,
if displacement occurs at the very end, the problem will be recreated,
since $Z$ will start in the same initial condition as in the usual
case without inflation.

  \section*{References}

\begin{enumerate}

\item G.Coughlan, W.Fischler, E.Kolb, S.Raby and G.Ross, {\it Phys.Lett.},
{\bf B131} (1983) 59; J.Ellis, D.V.Nanopoulos and M.Quiros, {\it Phys.Lett.},
{\bf B174} (1986) 176; G.German and G.G.Ross, {\it Phys.Lett.}, {\bf B172}
(1986) 305; O.Bertolami, {\it Phys.Lett} {\bf B209} (1988) 277;
R. de Carlos, J.A.Casas, F.Quevedo and E.Roulet, {\it Phys.Lett.} {\bf B318}
(1993) 447.

\item T.Banks, D.Kaplan and A.Nelson {\it Phys.Rev}, {\bf D49} (1994) 779.

\item L.Randall and S.Thomas, MIT preprint, LMU-TPW-94-17, hep-ph 9407248.

\item T.Banks, M.Berkooz and P.J.Steinhardt, preprint RU-94-92.

\item A.H.Guth, {\it Phys.Rev} {\bf D23} (1981) 347; For the review see
A.D.Linde, {\it Particle Physics and Inflationary Cosmology} (Harwood
Academic, Switzerland, 1990); E.W.Kolb and M.S.Turner, {\it The Early
Universe} (Addison-Wesley, Reading, MA, 1990);

\item G.Dvali, Q.Shafi and R.Schaefer, {\it Phys.Rev.Lett.} {\bf 73}
(1994) 1886.

\end{enumerate}
\end{document}